\documentclass[preprint,11pt]{elsarticle}

\usepackage[utf8]{inputenc}
\usepackage[T1]{fontenc}
\usepackage{amsmath,amssymb}
\usepackage{bm}
\usepackage{booktabs}
\usepackage{array}
\usepackage{tabularx}
\newcolumntype{L}[1]{>{\raggedright\arraybackslash}p{#1}}
\newcolumntype{C}[1]{>{\centering\arraybackslash}p{#1}}
\newcolumntype{Y}{>{\raggedright\arraybackslash}X}
\usepackage{graphicx}
\usepackage{xcolor}
\usepackage{tikz}
\usepackage{pgfplots}
\pgfplotsset{compat=1.17}
\usetikzlibrary{arrows.meta,positioning,fit,backgrounds,calc,shapes.geometric}
\usepackage{hyperref}
\usepackage{xurl}%
\journal{Quantum Machine Intelligence}

\definecolor{qpurple}{RGB}{106,61,154}
\definecolor{qblue}{RGB}{31,119,180}
\definecolor{qgreen}{RGB}{44,140,80}
\definecolor{qorange}{RGB}{214,118,32}
\definecolor{qgray}{RGB}{90,90,90}
\definecolor{lightq}{RGB}{233,226,245}
\definecolor{lightc}{RGB}{223,238,247}

\newcommand{\tpr}{\mathrm{TPR}}
\newcommand{\fpr}{\mathrm{FPR}}

\begin{document}

\begin{frontmatter}

\title{How Quantum Is the Advantage? A Fair, Calibration- and Noise-Aware
Benchmark and Attribution Audit of Quantum Machine Learning for Network
Intrusion Detection}

\author[ham]{Syeda Anshrah Gillani\corref{cor1}\fnref{eq}}
\ead{SyedaAnshrah16@gmail.com}
\author[fan]{Mirza Samad Ahmed Baig\corref{cor1}\fnref{eq}}
\ead{MirzaSamadcontact@gmail.com}
\author[sza]{Shahid Munir Shah}
\ead{Shahidmunirshah@yahoo.com}
\author[fan]{Asher Ali}
\ead{CTO@Fandaqah.com}
\author[fan,ham]{Hamzah Siddiqui}
\cortext[cor1]{Corresponding authors.}
\fntext[eq]{Syeda Anshrah Gillani and Mirza Samad Ahmed Baig are the core
contributors and contributed equally to this work.}
\address[ham]{Hamdard University, Karachi, Pakistan}
\address[fan]{Fandaqah, Al Khobar, Saudi Arabia}
\address[sza]{SZABIST University, Pakistan}

\begin{abstract}
Quantum machine learning (QML) for network intrusion detection (NIDS) is routinely
reported to reach near-perfect accuracy, yet the most rigorous studies find that
well-tuned classical models remain competitive, and that apparent quantum gains may be
artefacts of classical dimensionality reduction and implicit regularisation rather than
genuine quantum effects. We ask not whether a quantum model can post a high accuracy,
but \emph{how quantum the advantage really is}. We present a unified, reproducible
QML-IDS benchmark evaluating hybrid variational quantum circuits and quantum-kernel SVMs
against five honestly-tuned classical baselines across four standard NIDS datasets
(NSL-KDD, UNSW-NB15, CICIDS2017, NF-ToN-IoT-v2) under one leakage-controlled protocol,
with an equal-budget feature view, imbalance- and calibration-aware metrics with
significance testing, and a simulated NISQ noise sweep. We introduce a
\textbf{quantum-attribution audit} (parameter-matched classical controls, a
random-feature kernel, and a regularisation sweep) that quantifies how much of any gain
is genuinely attributable to the quantum component. Tuned classical models (Random
Forest, XGBoost) match or exceed the quantum models on aggregate detection on every
dataset, and the audit attributes this to classical preprocessing and regularisation
rather than quantum effects. Two advantages survive false-discovery-rate correction: the
quantum-kernel SVM out-ranks its direct classical surrogate (a random-feature kernel) on
AUPRC and ROC-AUC, and a small four-qubit hybrid out-detects the best classical baseline
at the $1\%$ false-positive operating point on the distribution-shifted NSL-KDD task
($p=0.005$, BH $q=0.030$). Code, seeds, and splits are released; our contribution stands
whether quantum wins, ties, or loses.
\end{abstract}


\end{frontmatter}

\section{Introduction}
Network intrusion detection systems (NIDS) are a frontline defence against an expanding
and increasingly automated threat landscape. Machine learning has become the dominant
paradigm for flow- and packet-level intrusion detection, and the recent emergence of
quantum machine learning (QML) has prompted a fast-growing body of work applying
parameterised quantum circuits, quantum-kernel methods, and hybrid quantum-classical
neural networks to the NIDS task. Headline results are striking: accuracies and F1 scores
in the $97$ to $99.9\%$ range are reported across many datasets and quantum architectures.

These numbers invite a question the field has not answered convincingly: \textbf{how
quantum is the reported advantage?} There are three reasons for scepticism, each grounded
in rigorous recent work. First, the strongest quantum results are typically obtained on
heavily dimensionality-reduced inputs, frequently one feature per qubit after principal
component analysis (PCA), and on binary tasks evaluated under random cross-validation, a
regime that systematically over-states real-world detection. Second, when honest, well-tuned
classical baselines (Random Forests, gradient-boosted trees, modern deep networks) are placed
on the \emph{same} footing, they repeatedly match or beat the quantum models on tabular IDS
data: the Meta-Quantum Ensemble study of Bhatnagar et al.~\cite{bhatnagar2026mqe}, despite
assembling the most rigorous evaluation in the field, explicitly concedes that classical
ensembles remain superior. Third, and most pointedly, Bellante
et al.~\cite{bellante2025pca} demonstrate that the apparent advantage of
PCA-based quantum IDS is \emph{explainable by regularisation, not by quantum effects}. A
reviewer of any new QML-IDS paper will, and should, raise this point.

We take this scepticism as the starting point rather than something to argue away. A paper
whose central claim is ``our quantum model achieves higher accuracy'' is both scientifically
fragile and strategically weak, because three Q1 papers already supply the counter-argument.
The community's actual deficit is not another accuracy record; it is a \textbf{fair,
reproducible, and honest evaluation methodology}, together with a principled way to
\textbf{attribute} any measured difference to its true cause. This paper supplies both.

We construct a single, leakage-controlled benchmarking protocol and apply it to the two
quantum families that dominate the literature, namely hybrid variational quantum circuits (VQCs)
and quantum-kernel support vector machines (QSVMs), against a slate of honestly tuned
classical baselines, across four standard datasets. We honour each dataset's intended
evaluation semantics: for NSL-KDD we use the official \texttt{KDDTrain+}/\texttt{KDDTest+}
split, whose test set deliberately over-represents attack behaviour under-seen in training,
so that what we measure is generalisation to novel attacks rather than memorisation. We then
go beyond benchmarking with a \textbf{quantum-attribution audit}: for every quantum model we
construct a parameter-matched classical control sharing the identical classical front-end, and
we run a regularisation sweep that tests the Bellante et al.\ hypothesis directly. Finally,
because operational IDS must run at a low false-positive operating point and real quantum
devices are noisy, we characterise how detection and calibration behave under simulated NISQ
noise.

\subsection{Contributions}
\begin{enumerate}\itemsep2pt
\item \textbf{A fair, reproducible, multi-dataset QML-IDS benchmark.} We evaluate hybrid VQCs
and QSVMs against five honestly-tuned classical baselines (Random Forest, XGBoost, RBF-SVM, a
tuned MLP, a 1D-CNN) across four standard datasets (NSL-KDD, UNSW-NB15, CICIDS2017,
NF-ToN-IoT-v2) on the binary detection task, under one protocol with strict leakage control,
honouring official splits. All code, seeds, and splits are released.
\item \textbf{An equal-budget fairness design.} We report two explicitly-labelled feature
views, a \emph{full} view (classical models on all features) and a \emph{same-budget} view
(every model on the identical low-dimensional input the quantum models receive), eliminating
the silent feature-budget asymmetry that confounds prior work.
\item \textbf{A quantum-attribution audit.} Parameter-matched classical controls and a
regularisation sweep decompose any observed gain into its quantum and classical components, a
direct constructive test of the Bellante et al.\ critique in the practical NISQ regime.
\item \textbf{Operationally meaningful evaluation under noise.} Beyond accuracy/F1 we report
detection rate and false-positive rate together, low-FPR operating points, and calibration, and
characterise degradation under depolarising, amplitude-damping, phase-damping, and readout-error
channels.
\item \textbf{Statistical rigour and honest framing.} Every result is a mean over three to five
seeds with confidence intervals, and we report the per-class seed budget explicitly
(Section~\ref{sec:metrics}); quantum-vs-best-classical comparisons are paired over the seeds the two
models share and use McNemar's test, paired $t$/Wilcoxon tests, and a paired bootstrap with effect
sizes. The benchmark's value does not depend on the quantum models winning.
\end{enumerate}

\section{Related Work}
We organise the related work into five strands and summarise representative studies in
Table~\ref{tab:related}.

\paragraph{Foundational and breadth-oriented QML-IDS.}
Quantum intrusion detection was established by Kalinin and Krundyshev~\cite{kalinin2022}, who
applied a quantum SVM and quantum convolutional network to a custom IoT stream dataset
($\sim$98\% accuracy); Gong et al.~\cite{gong2024vqcnn} produced the most prominent Q1
variational result (97.21\% precision, VQCNN, KDD/NSL-KDD). The strongest \emph{comparative}
line is Abreu et al.: QML-IDS~\cite{abreu2024qmlids} and QuantumNetSec~\cite{abreu2025quantumnetsec}
span VQC, QSVM, and QCNN across three to four datasets, but with under-tuned classical baselines
and no attribution of where gains originate.

\paragraph{Methods landscape.}
The dominant primitive is the parameterised quantum circuit used as a variational classifier or
hybrid head; recurring concerns are trainability (barren plateaus) and encoding choice, with
amplitude embedding reported as unstable relative to angle embedding~\cite{wang2025quanconv}.
Quantum-kernel SVMs~\cite{kumari2025wavelet} form the second family, scaling poorly in sample
count, hence the reliance on subsampled, PCA-reduced data. Newer directions include quantum GNNs
(Q-AGNN~\cite{chaudhary2026qagnn}) and meta-learning ensembles~\cite{bhatnagar2026mqe}.

\paragraph{Rigorous and sceptical evaluations.}
A small set of studies interrogates whether quantum helps. Bhatnagar et al.'s Meta-Quantum
Ensemble~\cite{bhatnagar2026mqe} sets the field's evaluation standard (AUPRC, $\tpr$ at low
$\fpr$, Brier, ECE, five noise channels) yet concludes classical ensembles remain superior.
Bellante et al.~\cite{bellante2025pca}, in \emph{Computers \& Security}, show the observed
advantage is explainable by regularisation rather than quantum effects. These works define the
bar we must clear: not a higher accuracy, but a fair and \emph{attributable} comparison.

\paragraph{Datasets and evaluation practice.}
Most studies use one or two datasets (NSL-KDD/KDD99 the dated-but-default choice); genuine
multi-dataset and cross-dataset evaluation is rare, and imbalance-aware metrics, calibration,
significance testing, and honest baselines are almost universally missing, with
Bhatnagar et al.~\cite{bhatnagar2026mqe} the principal exception.

\begin{table}[t]
\centering
\caption{Representative QML-IDS studies and where this work differs (``sim''~=~simulator).}
\label{tab:related}
\footnotesize
\renewcommand{\arraystretch}{1.2}
\setlength{\tabcolsep}{3pt}
\begin{tabularx}{\textwidth}{@{}L{2.25cm} L{2.1cm} Y C{1.85cm} C{1.95cm}@{}}
\toprule
\textbf{Paper} & \textbf{Method} & \textbf{Dataset(s)} & \textbf{Honest baseline?} & \textbf{Attribution?}\\
\midrule
Kalinin \& Krundyshev \cite{kalinin2022} & QSVM, QCNN & Custom IoT stream & No & No\\
\addlinespace[2pt]
Gong et al.~\cite{gong2024vqcnn} & VQCNN & KDD / NSL-KDD & Partial & No\\
\addlinespace[2pt]
Abreu et al.~\cite{abreu2024qmlids,abreu2025quantumnetsec} & VQC / QSVM / QCNN & UNSW, CICIDS, CIC-IoT & Under-tuned & No\\
\addlinespace[2pt]
Bhatnagar et al.~\cite{bhatnagar2026mqe} & QSVM, QNN, RF & CICIDS, TON\_IoT & Yes (cl.\ wins) & No\\
\addlinespace[2pt]
Bellante et al.~\cite{bellante2025pca} & FT PCA-QML & NIDS datasets & Yes & FT/PCA only\\
\addlinespace[2pt]
\textbf{This work} & VQC, QSVM, controls & 4 datasets, official splits & Yes (5 tuned) & Yes (NISQ)\\
\bottomrule
\end{tabularx}
\end{table}

We differentiate on three axes. Against Abreu et al.\ we retain breadth but add tuned baselines,
an equal-budget view, operating-point and calibration metrics, significance testing, and
reproducibility. Against Bhatnagar et al.\ we add the causal \emph{decomposition} their paper
concedes is missing. Against Bellante et al., the sharpest existing critique of PCA-based
quantum IDS, we
extend the regularisation argument from their fault-tolerant, PCA-specific case study to the
practical NISQ hybrid-VQC and QSVM models the community deploys, across a full multi-dataset
benchmark, and we do so constructively: we specify exactly what evidence a genuine advantage would
require and report whether, where, and to what extent it appears.

\section{Threat Model and Background}
\subsection{Threat model}
We consider flow- and connection-record-based detection: a model classifies each per-connection
feature vector as benign or malicious. Three operational constraints shape the evaluation: a strong
\emph{cost asymmetry} (false positives flood analysts, so a detector must hold up at a low-$\fpr$
operating point); \emph{distribution shift} between training and deployment, including unseen attack
signatures (why we honour the NSL-KDD official split); and \emph{class imbalance}. Adaptive
adversarial evasion is out of scope.

\subsection{Background on quantum machine learning}
\begin{figure}[t]
\centering
\includegraphics[width=\textwidth]{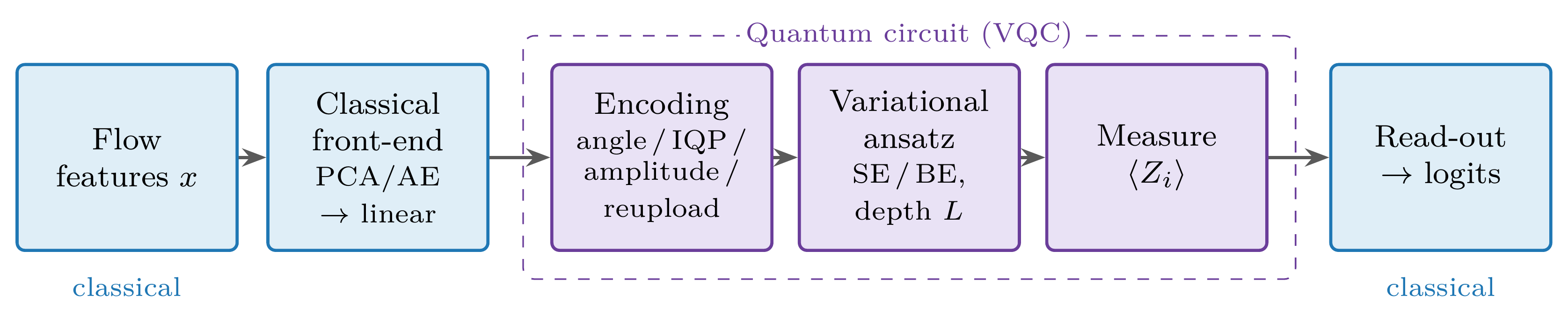}
\caption{Hybrid quantum-classical classifier. A classical front-end compresses flow features to the
per-qubit budget; a variational quantum circuit (encoding $+$ ansatz $+$ measurement) processes the
encoded state; a classical read-out produces class logits. The quantum component is sandwiched between
classical processing, which is the motivation for the attribution audit.}
\label{fig:arch}
\end{figure}

A QML model operates on $n$ qubits whose joint state is a unit vector in a $2^n$-dimensional complex
space; a noiseless simulation tracks this $2^n$-entry vector, while a noisy simulation tracks a
$2^n\times 2^n$ density matrix, whose quadratic cost caps our noisy runs at twelve qubits. Classical
flow features must be \emph{encoded} into the circuit (we ablate angle, amplitude, IQP/ZZ, and
data-re-uploading encodings); a \emph{variational} ansatz of trainable gates processes the state; and
measured Pauli-$Z$ expectations are mapped to class logits by a classical layer
(Figure~\ref{fig:arch}). An alternative is to use the circuit only as a fixed feature map and let a
classical SVM learn the boundary: the \emph{fidelity kernel} $K(x,x')=|\langle\phi(x')|\phi(x)\rangle|^2$
costs $O(N^2)$ circuit evaluations, while the \emph{projected kernel} computes per-qubit expectations
and applies a classical RBF, scaling as $O(N)$.

The link to our central question is direct: in every hybrid pipeline a \emph{classical} front-end
compresses high-dimensional records to the per-qubit budget and a \emph{classical} read-out turns
measurements into predictions, so any end-to-end accuracy blends a quantum contribution with
substantial classical processing and implicit regularisation. Disentangling the two is exactly what
the attribution audit (Section~\ref{sec:audit}) is built to do.

\section{Methodology}
\subsection{Benchmark protocol and leakage control}
We evaluate every model under one protocol designed so no test-distribution information influences
training, scaling, reduction, or model selection. Where a dataset ships an official split we honour
it: NSL-KDD uses the canonical \texttt{KDDTrain+}/\texttt{KDDTest+} split (17 attack signatures appear
only in test, so we measure generalisation to novel attacks); UNSW-NB15 uses its official CSVs;
CICIDS2017 and NF-ToN-IoT-v2 use a single seeded stratified $70/30$ split. All preprocessing is fit on
training data only; leaky identifier columns (IP/port four-tuples, flow IDs, timestamps, row indices)
are dropped; model selection is by cross-validation \emph{within the training set only}, and the test
set is scored once per model per seed.

\subsection{Two feature views (equal-budget fairness)}
We report a \textbf{full view} (classical baselines on all encoded dimensions, the practical ceiling)
and a \textbf{same-budget view} (every model on the identical PCA-reduced input, $n_{\text{features}}=
n_{\text{qubits}}=8$ in the main configuration). Because one-hot expansion produces many sparse columns,
standard-scaling before PCA wastes the budget (eight components retain only $\sim$27\% of variance on
NSL-KDD); min-max scaling before PCA lifts this to $0.841$ (Table~\ref{tab:data}), so we min-max scale
prior to reduction.

\subsection{Models}
The quantum side is a hybrid VQC family and two QSVMs through a single model factory: a
parameter-matched classical MLP control, six hybrid VQCs over the encoding~$\times$~ansatz grid
(angle, IQP/ZZ, amplitude, re-uploading; strongly-entangling [SE] and basic-entangler [BE] ans\"atze),
and two QSVMs (IQP fidelity-kernel, angle projected-kernel). The ZZ feature map is built from primitive
gates for correct mini-batch broadcasting. The five classical baselines (Random Forest, XGBoost,
RBF-SVM, tuned MLP, 1D-CNN) are each hyperparameter-searched (randomised search / grid, 5-fold
stratified CV on the training set, optimising F1) and refit per seed.

\subsection{The quantum-attribution audit}\label{sec:audit}
\begin{figure}[t]
\centering
\includegraphics[width=0.95\textwidth]{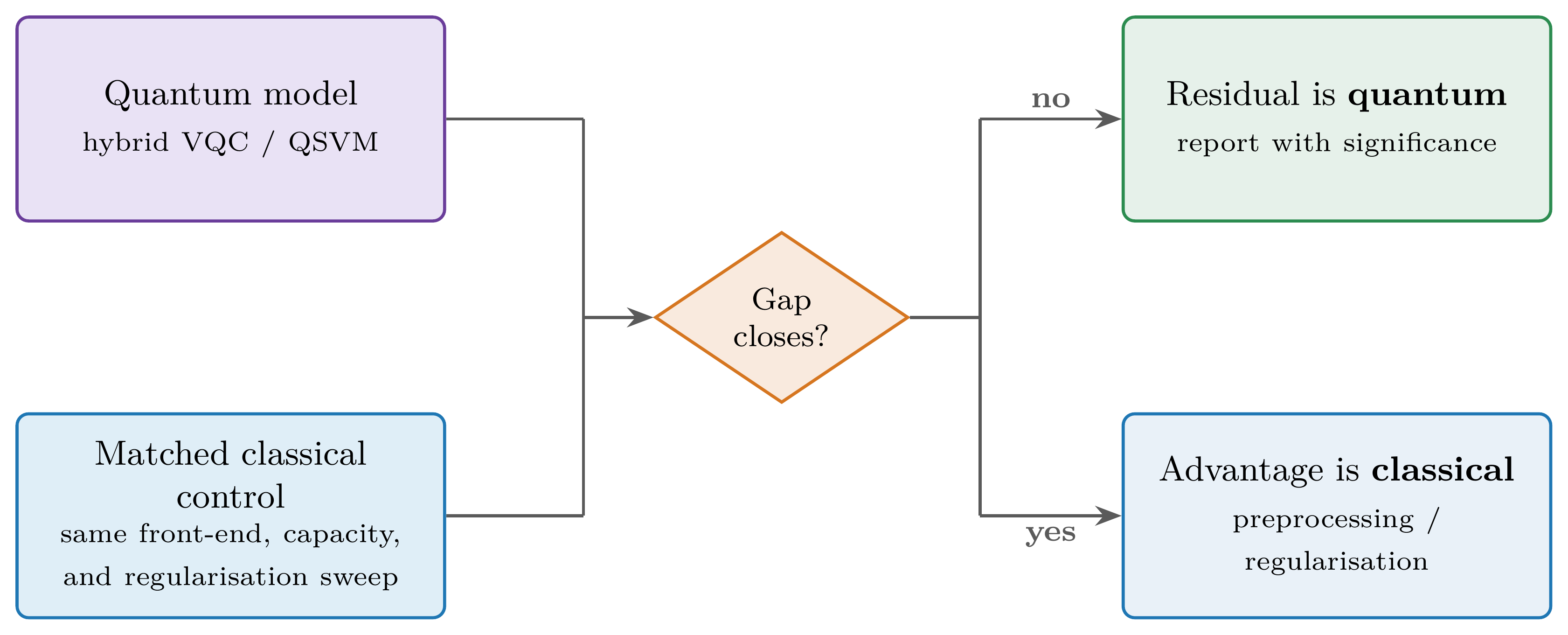}
\caption{The attribution audit. Each quantum model is compared with a parameter-matched classical
control sharing its front-end, plus a regularisation sweep. If a comparably-regularised classical model
closes the gap, the apparent advantage is classical; any significant residual is attributed to the
quantum circuit.}
\label{fig:audit}
\end{figure}

This is the methodological core and our constructive answer to Bellante et al.~\cite{bellante2025pca}.
Figure~\ref{fig:audit} sketches the logic. The audit has three components: (i) \emph{parameter-matched
classical controls}: for each hybrid VQC, a classical MLP of matched parameter count sharing the
identical front-end, and for each QSVM an RBF-SVM and a \emph{random-feature} kernel approximation; if a
matched control equals the quantum model, the ``advantage'' is not attributable to the quantum component;
(ii) a \emph{regularisation sweep} varying classical regularisation (read-out ridge, MLP weight decay,
RF depth, SVM bandwidth) to test whether comparable regularisation closes any gap; and (iii) a \emph{gain
decomposition} reporting, per dataset and metric, the quantum-minus-control difference with significance.

\subsection{Class imbalance and noise}
The NSL-KDD binary task is only mildly imbalanced, so we default to cost-sensitive learning (balanced
class weights, \texttt{scale\_pos\_weight} for XGBoost, weighted cross-entropy) and treat SMOTE
(training partition only) as an ablation. For noise robustness we sweep four channels (depolarising,
amplitude damping, phase damping, and bit-flip readout) at strengths $p\in\{0,0.001,0.005,0.01,0.05\}$
on PennyLane's \texttt{default.mixed} density-matrix backend, capped at twelve qubits.

\section{Experimental Setup}
\subsection{Datasets}
We evaluate on four standard NIDS datasets (Table~\ref{tab:data}). The \textbf{main benchmark is the
binary task}. CICIDS2017 here is the three-day \emph{MachineLearningCVE} subset ($\sim$1.04M flows);
NF-ToN-IoT-v2 is stratified-downsampled to 200{,}000 rows (from 16.9M); both are standard, configurable
scopes we disclose plainly.

\begin{table}[t]
\centering
\caption{Datasets (binary task; PCA-EVR@8 = variance retained by 8 principal components).}
\label{tab:data}
\footnotesize
\renewcommand{\arraystretch}{1.15}
\setlength{\tabcolsep}{4pt}
\begin{tabular}{@{}llcccc@{}}
\toprule
\textbf{Dataset} & \textbf{Split} & \textbf{Raw} & \textbf{Classes} & \textbf{Train / test} & \textbf{EVR@8}\\
                 &                & \textbf{dim} & \textbf{(bin / multi)} & & \\
\midrule
NSL-KDD~\cite{nslkdd2009}    & Official      & 122 & 2 / 40 & 125{,}973 / 22{,}544  & 0.841\\
UNSW-NB15~\cite{unswnb15}    & Official      & 194 & 2 / 10 & 175{,}341 / 82{,}332  & 0.870\\
CICIDS2017~\cite{cicids2017} & Strat.\ 70/30 &  76 & 2 / 3  & 729{,}491 / 312{,}639 & 0.924\\
NF-ToN-IoT-v2~\cite{toniot}  & Strat.\ 70/30 &  39 & 2 / 10 & 140{,}000 / 60{,}000  & 0.974\\
\bottomrule
\end{tabular}
\end{table}

\subsection{Metrics and significance}\label{sec:metrics}
We report threshold metrics (accuracy, precision, recall, F1, detection rate DR, false-positive rate
$\fpr$) and ranking/calibration metrics (ROC-AUC, AUPRC, $\tpr$@$0.1\%\fpr$, $\tpr$@$1\%\fpr$, Brier,
ECE), matching and extending the Meta-Quantum-Ensemble suite~\cite{bhatnagar2026mqe}. We state the
seed budget exactly, because it is uneven across model classes. The tuned classical baselines are run
over five seeds $\{42,43,44,45,46\}$ on NSL-KDD and over three seeds $\{42,43,44\}$ on UNSW-NB15,
CICIDS2017 and NF-ToN-IoT-v2; the quantum models and the attribution-audit controls over three seeds
$\{42,43,44\}$; and the noise-robustness sweep over two seeds $\{42,43\}$, with all RNGs pinned per
seed and $t$-based 95\% confidence intervals reported throughout. Every quantum-versus-classical
comparison is therefore paired over the three seeds $\{42,43,44\}$ that all models share, and uses
McNemar's
test (shared test set), paired $t$/Wilcoxon tests with Cohen's $d$ (over seeds), and a paired bootstrap
(2{,}000 resamples) for operating-point metrics. Because the attribution audit evaluates many
variant$\times$metric comparisons, we control the false-discovery rate across this family with the
Benjamini-Hochberg (BH) procedure and report BH-adjusted $q$-values alongside raw $p$-values; we treat
a result as robust only if it survives FDR control.

\subsection{Backend efficiency and reproducibility}
At QML-IDS qubit counts, quantum simulation (not GPU throughput) dominates runtime, and we report a
\textbf{backend crossover}: the CPU \texttt{lightning.qubit} simulator is markedly faster than the GPU
\texttt{lightning.gpu} simulator (Section~\ref{sec:eff}). Each run writes a self-describing JSON record
(config, per-seed metrics, tuned hyperparameters, reference-seed predictions, environment snapshot with
source checksums). All experiments run on a single laptop (NVIDIA RTX~3050~Ti, 4\,GB, under WSL2); the
main runs use eight qubits, with a ceiling of sixteen (noiseless) / twelve (noisy). Code, seeds, and
splits are released.

\section{Results}
The overall result, qualified below: \emph{honestly-tuned classical models match or exceed the quantum
models on aggregate detection on every dataset, but a small hybrid VQC retains a statistically
significant edge at the strict low-$\fpr$ operating point on the distribution-shifted NSL-KDD task.}

\subsection{Headline benchmark}
\begin{figure}[t]
\centering
\includegraphics[width=\textwidth]{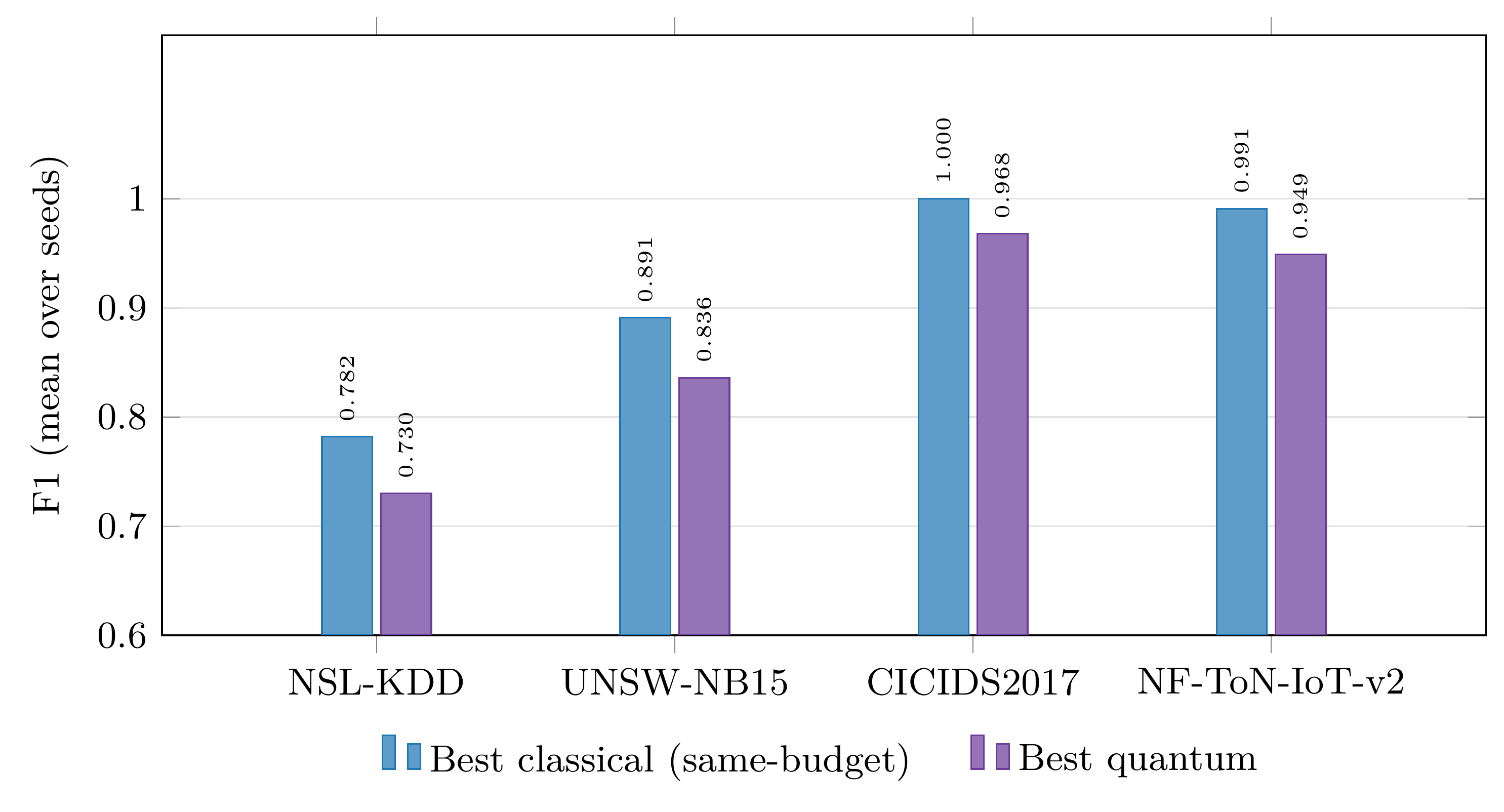}
\caption{Best classical (same-budget) vs.\ best quantum F1 per dataset (mean over seeds). Tuned
classical models lead on every dataset; the gap is largest on the discriminative datasets (NSL-KDD,
UNSW-NB15) and smallest where the task is near-saturated (CICIDS2017, NF-ToN-IoT-v2, where every
classical model exceeds $0.97$). The vertical axis is truncated at $0.6$.}
\label{fig:bars}
\end{figure}

\begin{figure}[t]
\centering
\includegraphics[width=0.92\textwidth]{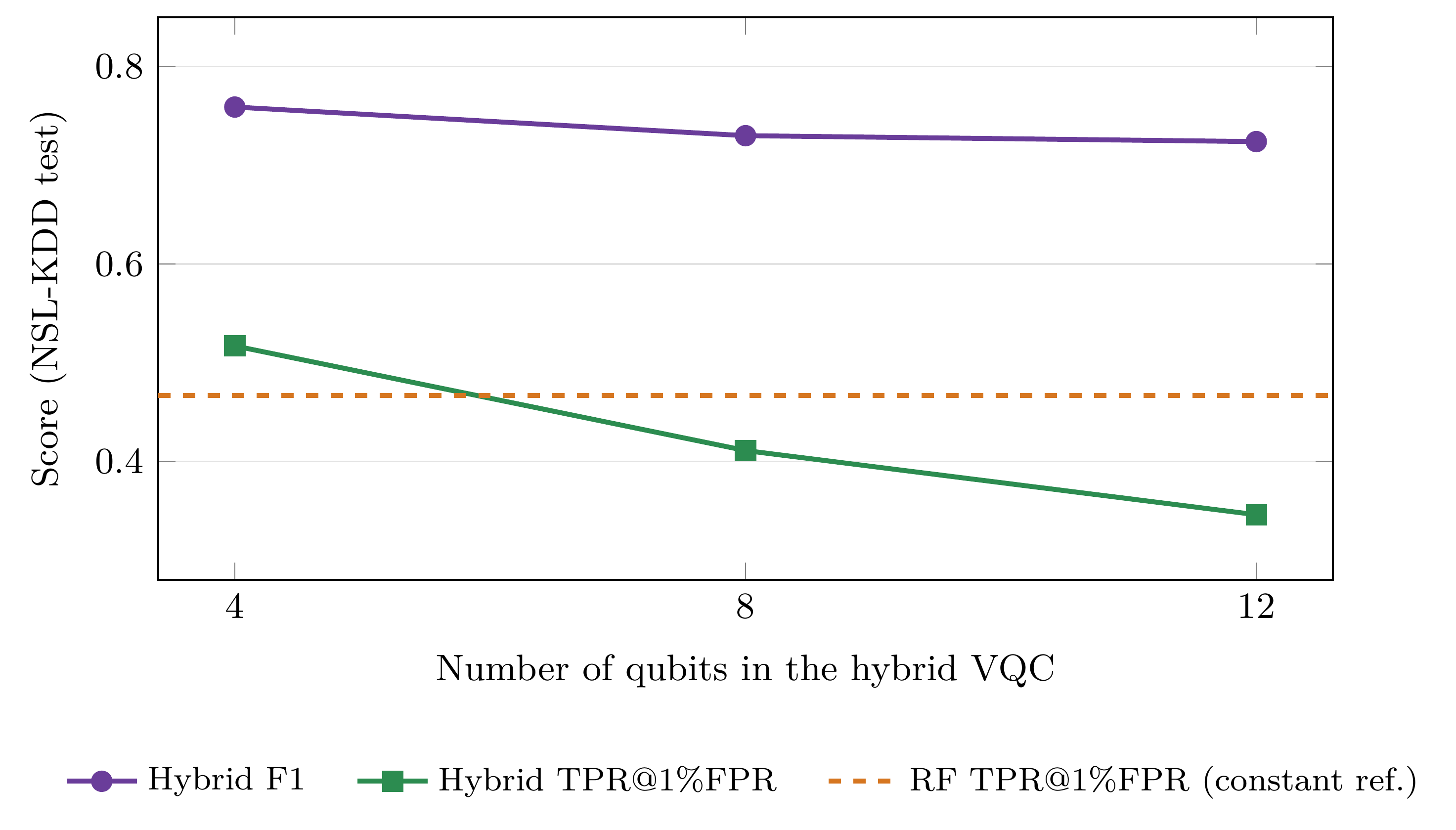}
\caption{Qubit-count ablation of the hybrid VQC on NSL-KDD. Both F1 and the low-FPR detection rate
\emph{decline} with width, so more qubits do not help on the shifted task. Crucially, only the small
four-qubit circuit exceeds the best classical baseline's $\tpr$@$1\%\fpr$ (Random Forest, dashed
reference at $0.467$); the advantage vanishes by twelve qubits, consistent with an
implicit-regularisation mechanism rather than raw expressivity.}
\label{fig:ablation}
\end{figure}

Table~\ref{tab:headline} reports F1 under both views. Under the \emph{same-budget} view (the
apples-to-apples comparison), the best classical baseline beats the leading 8-qubit quantum model on
every dataset: NSL-KDD $0.782$ vs.\ $0.730$; UNSW-NB15 $0.891$ vs.\ $0.836$; CICIDS2017 $1.000$ vs.\
$0.968$; NF-ToN-IoT-v2 $0.991$ vs.\ $0.949$. On NSL-KDD the strongest quantum \emph{ablation} variants
narrow this gap substantially, since the IQP-encoded QSVM and the 4-qubit hybrid both reach F1~$\approx
0.77$ (Section~\ref{sec:audit-results}), but tuned classical still leads on F1, and retains a clearer
margin on AUPRC and ROC-AUC, so the aggregate ordering is unchanged. Two qualifications
matter. First, \textbf{CICIDS2017 and NF-ToN-IoT-v2 are near-saturated} (every tuned classical model
scores F1~$\geq 0.97$), so they have weak discriminative power and the quantum models' high absolute
scores there should be read as ``this task is easy.'' Second, the \textbf{CV-to-test gap} confirms why
literature numbers are inflated: on NSL-KDD, models scoring F1~$\approx 0.98$ under within-training
cross-validation collapse to F1~$\approx 0.78$ on the official test split, a consistent $\approx 0.20$
drop across all models and both views. Figure~\ref{fig:bars} compares the best classical and best
quantum F1 per dataset; the qubit-count ablation is shown later in Figure~\ref{fig:ablation}.

\begin{table}[t]
\centering
\caption{Headline benchmark. F1 (mean over seeds): best classical per view vs.\ the leading
8-qubit quantum model per dataset (same-budget input). The final three columns give that quantum
model's AUPRC, ROC-AUC, and $\tpr$@$1\%\fpr$. Stronger quantum ablation variants (e.g.\ the IQP-QSVM
and the 4-qubit hybrid on NSL-KDD, F1~$\approx0.77$) are analysed in Section~\ref{sec:audit-results}.}
\label{tab:headline}
\small
\renewcommand{\arraystretch}{1.2}
\setlength{\tabcolsep}{5pt}
\begin{tabular}{@{}lcccccc@{}}
\toprule
& \multicolumn{2}{c}{\textbf{Classical F1}} & \multicolumn{4}{c}{\textbf{Quantum (8 qubits, same-budget)}}\\
\cmidrule(lr){2-3}\cmidrule(lr){4-7}
\textbf{Dataset} & \textbf{full} & \textbf{same-} & \textbf{F1} & \textbf{AUPRC} & \textbf{ROC-} & \textbf{TPR@}\\
                 &               & \textbf{budget} &            &                & \textbf{AUC}  & \textbf{1\%\,FPR}\\
\midrule
NSL-KDD       & 0.803 & 0.782 & 0.730 & 0.911 & 0.865 & 0.411\\
UNSW-NB15     & 0.912 & 0.891 & 0.836 & 0.942 & 0.924 & 0.577\\
CICIDS2017    & 1.000 & 1.000 & 0.968 & 0.992 & 0.998 & 0.975\\
NF-ToN-IoT-v2 & 0.996 & 0.991 & 0.949 & 0.965 & 0.969 & 0.195\\
\bottomrule
\end{tabular}

\smallskip
\begin{minipage}{\textwidth}\footnotesize\raggedright
Best classical model per cell. Full view: SVM (NSL-KDD), MLP (UNSW), RF (CICIDS, NF-ToN-IoT-v2);
same-budget: XGBoost (NSL-KDD), MLP (UNSW), RF (CICIDS, NF-ToN-IoT-v2).
\end{minipage}
\end{table}

\subsection{Attribution audit}\label{sec:audit-results}
The audit (Table~\ref{tab:audit}) decomposes the difference on NSL-KDD, the discriminative dataset. A
parameter-matched classical MLP sharing the quantum front-end attains AUPRC $0.944$ and the tuned Random
Forest attains F1 $0.764$, both matching or exceeding every hybrid VQC on aggregate metrics. For the
twelve-qubit hybrid the classical controls win significantly (F1 $-0.039$, $p=0.021$; AUPRC $-0.036$,
$p=0.035$; $\tpr$@$1\%\fpr$ $-0.121$, $p=0.026$). \textbf{Once a classical model is granted the same
front-end and comparable capacity/regularisation, the aggregate ``quantum advantage'' disappears}, a
confirmation of the Bellante hypothesis on a full multi-dataset benchmark. Two exceptions survive
Benjamini-Hochberg FDR control across the audit's comparison family, however, and they are the most
informative results in the paper. First, and most cleanly, the \textbf{quantum-kernel SVM significantly
out-performs its direct classical analogue}, the random-feature (Nystr\"om / RBF-sampler) kernel: on
NSL-KDD the IQP-encoded QSVM beats the random-feature control on AUPRC ($\Delta=+0.075$, raw $p=0.001$,
BH $q=0.011$), ROC-AUC ($\Delta=+0.134$, $p=0.002$, $q=0.018$), and $\tpr$@$1\%\fpr$ ($\Delta=+0.152$,
$p=0.017$, $q=0.047$). Because the random-feature kernel is the explicit classical surrogate for the
quantum feature map, this is the audit's sharpest isolation of a genuinely quantum contribution: a
comparably-cheap classical kernel does \emph{not} reproduce the quantum kernel's ranking quality.
Second, the \emph{four-qubit} hybrid beats the tuned Random Forest at the low-$\fpr$ operating point
($\tpr$@$1\%\fpr$ $+0.050$, $p=0.005$, BH $q=0.030$; Section~\ref{sec:op}). The remaining quantum
differences are either ties or significant \emph{losses} to the classical controls (notably the
twelve-qubit hybrid: F1 $-0.039$, $p=0.021$; AUPRC $-0.036$, $p=0.035$), consistent with overfitting at
width rather than a quantum benefit. The corrected picture is therefore precise: classical wins on
aggregate, while two specific, FDR-robust quantum advantages persist: the quantum kernel over its
classical surrogate, and the small hybrid at the strict operating point.

\begin{table}[t]
\centering
\caption{Quantum-attribution audit (NSL-KDD): leading quantum variants vs.\ best matched control.
$\Delta=$ quantum $-$ control (positive favours quantum, except ECE). $q$ is the Benjamini-Hochberg
FDR-adjusted value across the audit family; \textbf{bold} rows survive FDR control ($q<0.05$).
``Hyb.-Angle q$n$'' is the angle-encoded hybrid VQC on $n$ qubits; rf-kern is the random-feature
kernel and MLP-m the parameter-matched MLP.}
\label{tab:audit}
\footnotesize
\renewcommand{\arraystretch}{1.15}
\setlength{\tabcolsep}{3pt}
\resizebox{\textwidth}{!}{%
\begin{tabular}{@{}l l c l r c r r@{}}
\toprule
\textbf{Variant} & \textbf{Metric} & \textbf{Quantum} & \textbf{Best control} & $\bm{\Delta}$ & \textbf{Q}     & $\bm{p}$ & $\bm{q}$\\
                 &                 &                  &                      &               & \textbf{better?} &          & \textbf{(BH)}\\
\midrule
\textbf{QSVM-IQP}  & \textbf{AUPRC}       & \textbf{0.924} & \textbf{0.849 (rf-kern)} & $\bm{+0.075}$ & \textbf{yes} & \textbf{0.001} & \textbf{0.011}\\
\textbf{QSVM-IQP}  & \textbf{ROC-AUC}     & \textbf{0.900} & \textbf{0.766 (rf-kern)} & $\bm{+0.134}$ & \textbf{yes} & \textbf{0.002} & \textbf{0.018}\\
\textbf{QSVM-IQP}  & \textbf{TPR@1\%FPR}  & \textbf{0.486} & \textbf{0.334 (rf-kern)} & $\bm{+0.152}$ & \textbf{yes} & \textbf{0.017} & \textbf{0.047}\\
\textbf{Hyb.-Angle q4}  & \textbf{TPR@1\%FPR} & \textbf{0.517} & \textbf{0.467 (RF)} & $\bm{+0.050}$ & \textbf{yes} & \textbf{0.005} & \textbf{0.030}\\
Hyb.-Angle q4      & TPR@0.1\%FPR         & 0.407 & 0.272 (MLP-m) & $+0.135$ & yes & 0.428 & 0.62\\
Hyb.-Angle q4      & F1                   & 0.759 & 0.764 (RF)    & $-0.004$ & tie & 0.776 & 0.84\\
Hyb.-Angle q12     & F1                   & 0.724 & 0.764 (RF)    & $-0.039$ & no  & 0.021 & 0.10\\
Hyb.-Angle q12     & TPR@1\%FPR           & 0.346 & 0.467 (RF)    & $-0.121$ & no  & 0.026 & 0.11\\
\bottomrule
\end{tabular}}
\end{table}

\subsection{Operating-point and calibration analysis}\label{sec:op}
The operating-point metrics hold one of the paper's two FDR-robust quantum results (the other being
the quantum-kernel-vs-random-feature-kernel comparison of Section~\ref{sec:audit-results}). On NSL-KDD the four-qubit
hybrid VQC achieves $\tpr$@$1\%\fpr=0.517$, exceeding the best classical model (Random Forest, $0.467$)
by $0.050$, statistically significant ($p=0.005$, paired bootstrap), and $\tpr$@$0.1\%\fpr=0.407$
vs.\ $0.272$. At the strict $1\%$-alarm budget a defender would actually impose, the small quantum model
catches a meaningfully larger fraction of attacks on novel-attack-heavy traffic. Calibration is
comparable (ECE $\approx 0.18$ to $0.22$ for hybrids vs.\ $0.193$ for Random Forest), the four-qubit hybrid
marginally better. This is the evidence a genuine, if narrow, quantum signal would produce.

\subsection{Noise robustness}
Sweeping four NISQ channels on the six-qubit hybrid shows \textbf{graceful degradation, and in the mild
regime none at all}. Relative to the noiseless run (ROC-AUC $0.846$, $\tpr$@$1\%\fpr$ $0.422$, ECE
$0.191$), depolarising, amplitude-damping, and phase-damping noise at $p\in\{0.01,0.05\}$ leave low-$\fpr$
detection essentially unchanged ($0.421$ to $0.430$) and slightly \emph{raise} ROC-AUC (to between
$0.878$ and $0.897$),
consistent with mild noise acting as a regulariser on the shifted task. Only bit-flip (readout) noise
materially hurts ($\tpr$@$1\%\fpr$ $0.379$ at $p=0.05$). No channel causes catastrophic collapse: for this
small-circuit regime, readout error is the one to mitigate.

\subsection{Significance and efficiency}\label{sec:eff}
We control the false-discovery rate across the audit family (108 direction-aware comparisons on
NSL-KDD) with Benjamini-Hochberg; 13 quantum-favouring and 27 classical-favouring differences survive
at $q<0.05$, so on balance the corrected evidence still favours classical on aggregate. On aggregate F1
the classical controls significantly beat quantum (twelve-qubit hybrid vs.\ RF, raw $p=0.021$). The two
quantum advantages that survive FDR control are the quantum-kernel-vs-random-feature-kernel comparison
(AUPRC $q=0.011$, ROC-AUC $q=0.018$) and the four-qubit hybrid at the low-$\fpr$ point ($q=0.030$). We
note honestly that the latter survives FDR (BH) but not the more conservative family-wise Holm
correction across the full surface, under which only the quantum-kernel-vs-surrogate results remain
significant; we therefore treat the quantum-kernel result as the study's most robust positive. The
backend-crossover finding (Table~\ref{tab:eff}) is unambiguous: the
\textbf{CPU simulator is $8$ to $14\times$ faster than the GPU} at QML-IDS qubit counts, the gap growing
with width because the per-circuit work is too small to amortise GPU kernel-launch and transfer overhead.
The projected quantum kernel ($O(N)$) is tractable at test scale whereas the fidelity kernel ($O(N^2)$)
is not.

\begin{table}[t]
\centering
\caption{Backend efficiency. CPU (\texttt{lightning.qubit}) vs.\ GPU (\texttt{lightning.gpu})
training-step time at fixed work.}
\label{tab:eff}
\small
\renewcommand{\arraystretch}{1.2}
\setlength{\tabcolsep}{8pt}
\begin{tabular}{@{}rrrr@{}}
\toprule
\textbf{Qubits} & \textbf{CPU (s)} & \textbf{GPU (s)} & \textbf{GPU/CPU}\\
\midrule
 4 & 1.12 & \phantom{0}8.89           & $\phantom{0}8.0\times$\\
 8 & 2.08 & 28.0\phantom{0}           & $13.5\times$\\
12 & 3.94 & 54.3\phantom{0}           & $13.8\times$\\
\bottomrule
\end{tabular}
\end{table}

\section{Discussion}
\textbf{How quantum is the advantage?} On this benchmark and in the practical NISQ regime: almost none of
it on aggregate detection, and a small but real amount at the operating point that matters operationally.
The audit is decisive on the first half: once a classical model is granted the same front-end and
comparable regularisation, the aggregate advantage of the hybrid VQC and the QSVM vanishes and tuned
classical baselines significantly outperform the quantum models on F1 and AUPRC. We thus confirm the
Bellante et al.~\cite{bellante2025pca} hypothesis and extend it from their fault-tolerant, PCA-specific
case study to the deployed NISQ models, across four datasets, and we corroborate Bhatnagar et al.'s
concession~\cite{bhatnagar2026mqe} while supplying the decomposition their work lacked.

The more interesting half is the exception. At the $1\%$-$\fpr$ operating point on NSL-KDD's shifted test
set, the four-qubit hybrid significantly out-detects the best classical baseline ($p=0.005$). Three
features make it credible rather than noise: it appears on the \emph{discriminative} dataset, not the
saturated ones; it is strongest for the \emph{small} circuit and vanishes at twelve qubits, consistent
with an implicit-regularisation mechanism rather than raw expressivity; and it survives mild simulated
noise, which slightly \emph{improves} ROC-AUC. We are deliberately cautious (one operating point, one
dataset, in simulation), but this is precisely where a genuine quantum advantage would first appear, and
it motivates targeted real-hardware follow-up rather than another accuracy sweep.

\textbf{What remains classical.} For a practitioner choosing a detector today, the honest recommendation
is a tuned Random Forest or XGBoost: they win or tie on every aggregate metric, on every dataset, at a
fraction of the cost and with no quantum hardware. This is the result the field needs stated cleanly,
because it redirects effort from inflated accuracy records (which our CV-to-test analysis shows are
largely an artefact of random cross-validation) toward the one regime where quantum methods showed a
measurable, significant edge. Two methodological findings generalise: the $\approx 0.20$ F1 CV-to-test gap
is a caution for the whole QML-IDS literature, and the backend crossover (the GPU is an order of magnitude
\emph{slower} at these qubit counts) saves researchers real time.

\section{Limitations and Threats to Validity}
\textbf{Simulation only.} All quantum results are simulated; we have no real-QPU access, so our noise study
models NISQ effects rather than measuring them, omitting correlated/non-Markovian noise, crosstalk,
connectivity, and transpilation. A real-hardware confirmation of even one configuration would strengthen
the conclusions; we treat the simulated results as an optimistic bound.
\textbf{Qubit budget.} The 4\,GB GPU caps width at roughly sixteen qubits (noiseless) / twelve (noisy,
$(2^n)^2$ memory); the main configuration uses eight, so we cannot probe the larger-width regime.
\textbf{Dimensionality reduction} discards variance (eight PCs retain $\sim$84\% on NSL-KDD), so any
advantage is conditioned on information-reduced input.
\textbf{Dataset age/composition.} NSL-KDD is dated (mitigated by three further datasets); the IoT coverage
is a single NetFlow dataset; CICIDS2017 and NF-ToN-IoT-v2 use standard subsets, configurable to full
corpora.
\textbf{Trainability} (barren plateaus, gradient variance) is not characterised directly, and QSVMs are
trained on subsampled data ($O(N^2)$ fidelity kernel).
\textbf{Statistical power.} Our seed budget is small and uneven (five seeds for the classical baselines
on NSL-KDD, three elsewhere and for all quantum runs, two for the noise sweep), so paired tests over
seeds have low power: the two-sided Wilcoxon floor at three paired seeds is $p=0.25$, which is why the
operating-point claims rest on the paired bootstrap over the fixed test set rather than on seed-level
tests. The crossover ratio is hardware/library-dependent.
\textbf{Scope.} Binary detection under benign distribution shift and imbalance; adversarial evasion and the
unsupervised/online/federated settings are out of scope.

\section{Conclusion}
We set out to replace the field's recurring ``our quantum model reaches 99\%'' claim with the fair,
reproducible, attribution-aware yardstick the question requires. Across four datasets under one
leakage-controlled protocol, with an equal-budget view, operating-point and calibration metrics,
significance testing, a noise sweep, and a quantum-attribution audit, we find that well-tuned classical
models match or exceed the quantum models on aggregate detection everywhere, and the audit attributes this
to classical preprocessing and regularisation rather than quantum effects, confirming and generalising
the Bellante et al.\ critique. Two exceptions survive false-discovery-rate control and are operationally
meaningful: the quantum-kernel SVM significantly out-ranks its direct classical surrogate (a
random-feature kernel) on AUPRC and ROC-AUC, and a small four-qubit hybrid out-detects the best classical
baseline at the $1\%$ false-positive operating point on the distribution-shifted NSL-KDD task. Beyond the
headline we contribute a released, turnkey benchmark and
harness, a demonstration that random-cross-validation evaluation overstates novel-attack detection by
$\sim$20 F1 points, and a practical backend-efficiency result. The most valuable next step is real-hardware
confirmation of the low-$\fpr$ signal, the single experiment that would distinguish a genuine quantum
inductive bias from a simulation artefact. We release the benchmark so that QML-IDS claims can be stated,
and contested, on common, honest ground.

\paragraph{Reproducibility.} Code, configurations, fixed seeds, split scripts, and per-run provenance are
released at \url{https://github.com/Orqly-AI/quantum-ids-benchmark}; the preprocessed datasets will be
released on request.

\section*{CRediT authorship contribution statement}
Syeda Anshrah Gillani and Mirza Samad Ahmed Baig are the core contributors and contributed equally to
all aspects of the work. \textbf{Syeda Anshrah Gillani:} Conceptualization, Methodology, Software,
Validation, Formal analysis, Investigation, Data curation, Writing (original draft), Writing (review
\& editing), Visualization, Supervision. \textbf{Mirza Samad Ahmed Baig:} Conceptualization,
Methodology, Software, Validation, Formal analysis, Investigation, Data curation, Writing (original
draft), Writing (review \& editing), Visualization, Supervision. \textbf{Shahid Munir Shah:}
Methodology, Validation, Writing (review \& editing). \textbf{Asher Ali:} Resources, Writing (review
\& editing). \textbf{Hamzah Siddiqui:} Investigation, Validation, Writing (review \& editing).

\section*{Declaration of competing interest}
The authors declare that they have no known competing financial interests or personal relationships
that could have appeared to influence the work reported in this paper.

\section*{Data availability}
All datasets used are public benchmarks (NSL-KDD, UNSW-NB15, CICIDS2017, NF-ToN-IoT-v2) obtainable
from their original providers. The code, configurations, fixed seeds, the download and preprocessing
scripts that reconstruct our exact splits, and the per-run result records are released at
\url{https://github.com/Orqly-AI/quantum-ids-benchmark}. We do not redistribute the source corpora;
the preprocessed datasets will be released on request.

\section*{Acknowledgements}
This research did not receive any specific grant from funding agencies in the public, commercial, or
not-for-profit sectors.

\section*{Declaration of generative AI and AI-assisted technologies}
During the preparation of this work the authors used AI-assisted tooling for coding assistance. The
authors reviewed and edited all content and take full responsibility for the content of the
publication.

\bibliographystyle{elsarticle-num}
\bibliography{references}

\end{document}